\documentclass[twocolumn,showpacs,prb,aps]{revtex4-1}%
\usepackage{amsfonts}
\usepackage{amsmath}
\usepackage{amssymb}
\usepackage{graphicx}
\usepackage{txfonts}%
\providecommand{\U}[1]{\protect\rule{.1in}{.1in}}

\begin{document}

\title{Strongly polarizing weakly coupled $^{13}$C nuclear spins with
optically pumped nitrogen-vacancy center}
\author{Ping Wang}
\affiliation{Hefei National Laboratory for Physics Sciences at Microscale and Department of
Modern Physics, University of Science and Technology of China, Hefei, Anhui
230026, China}
\affiliation{Beijing Computational Science Research Center, Beijing 100084, China}
\author{Bao Liu}
\affiliation{Beijing Computational Science Research Center, Beijing 100084, China}
\author{Wen Yang}
\email{wenyang@csrc.ac.cn}
\affiliation{Beijing Computational Science Research Center, Beijing 100084, China}

\date{\today}

\begin{abstract}
Enhancing the polarization of nuclear spins surrounding the nitrogen-vacancy
(NV) center in diamond has attracted widespread attention recently due to
its various applications. Here we present an analytical theory and
comprehensive understanding on how to optimize the dynamic nuclear
polarization by an optically pumped NV center near the ground state level
anticrossing. Our results not only provide a parameter-free explanation and
a clearly physics picture for the recently observed polarization dependence
on the magnetic field for strongly coupled $^{13}$C nuclei [H. J. Wang
\textit{et al}., Nat. Commun. 4, 1 (2013)], but also demonstrate the
possibility to strongly polarize weakly coupled $^{13}$C nuclei under weak
optical pumping and suitably chosen magnetic field. This allows sensitive
magnetic control of the $^{13}$C nuclear spin polarization for NMR
applications and significant suppression of the $^{13}$C nuclear spin noise
to prolong the NV spin coherence time.
\end{abstract}

\maketitle

\section*{Introduction}

The atomic nuclear spins are central elements for NMR and magnetic resonance
imaging \cite{AbragamBook1961} and promising candidates for storing and
manipulating long-lived quantum information \cite{KaneNature1998} due to
their long coherence time. However, the tiny magnetic moment of the nuclear
spins makes them completely random in thermal equilibrium, even in a strong
magnetic field and at low temperature. This poses severe limitations on
their applications. The dynamic nuclear polarization (DNP) technique can
bypass this limitation by transferring the electron spin polarization to the
nuclear spins via the hyperfine interaction (HFI), but efficient DNP is
usually prohibited at room temperature.

An exception is the negatively charged nitrogen-vacancy (NV) center \cite%
{GruberScience1997} in diamond, which has an optically polarizable spin-1
electronic ground state with a long coherence time \cite%
{BalasubramanianNatMater2009}, allowing DNP at room temperature \cite%
{HePRB1993,JacquesPRL2009}. This prospect has attracted widespread interest
due to its potential applications in room-temperature NMR, magnetic
resonance imaging and magnetometry \cite{MazeNature2008,TaylorNatPhys2008},
electron-nuclear hybrid quantum register \cite%
{ChildressScience2006,DuttScience2007,NeumannScience2010}, and electron spin
coherence protection by suppressing the nuclear spin noise \cite%
{ToganNature2011}. In addition to the remarkable success in coherently
driving spectrally resolved transitions to initialize, manipulate, and
readout up to three strongly coupled nuclear spins \cite%
{DuttScience2007,NeumannScience2008,JiangScience2009,WaldherrNature2014,TaminiauNatNano2014}%
, there are intense activities aiming to enhance the polarization of many
nuclear spins via dissipative spin transfer from the NV to the nuclear
spins. To overcome the large energy mismatch for resonant spin transfer,
various strategies have been explored, e.g., tuning the NV spin near the
excited state level anticrossing \cite%
{JacquesPRL2009,FischerPRL2013,FischerPRB2013,GaliPRB2009a,SmeltzerPRA2009,SteinerPRB2010,DreauPRB2012}
or ground state level anticrossing (GSLAC) \cite%
{HePRB1993,GaebelNature2006,WangNatCommun2013}, driving the NV-nuclear spins
into Hartman-Hahn resonance \cite{LondonPRL2013,LiuNanoscale2014} or
selectively driving certain spectrally resolved transitions between
hyperfine-mixed states under optical illumination \cite%
{AlvarezArxiv2014,KingArxiv2015}. Successful polarization of bulk nuclear
spins in diamond have dramatically enhanced the NMR signal by up to five
orders of magnitudes \cite{FischerPRL2013,KingArxiv2015} and significantly
prolonged the NV spin coherence time \cite{LondonPRL2013,LiuNanoscale2014}.

In particular, near NV excited state level anticrossing, almost complete
polarization has been achieved for $^{15}$N (or $^{14}$N) and $^{13}$C
nuclei in the first shell of the vacancy \cite%
{FuchsPRL2008,JacquesPRL2009,SmeltzerPRA2009,SteinerPRB2010,DreauPRB2012}.
Recently, Wang \textit{et al. }\cite{WangNatCommun2013} exploited the GSLAC
to achieve near complete polarization of first-shell $^{13}$C nuclei and
revealed multiple polarization sign reversals over a narrow range (a few mT)
of magnetic field. This behavior has been attributed to the anisotropic HFI
and could allow sensitive magnetic control of $^{13}$C nuclear spin
polarization, but a clear understanding remains absent. Furthermore, in most
of the existing works, only a few strongly coupled nuclear spins (HFI $\gg
200$ kHz) are significantly polarized via direct spin transfer from the NV
center, while many weakly coupled nuclear spins are only slightly polarized
via nuclear spin diffusion. Enhancing the polarization of these weakly
coupled nuclear spins could further improve NMR and magnetic resonance
imaging \cite{FischerPRL2013,AlvarezArxiv2014,KingArxiv2015} and prolong the
NV spin coherence time \cite{LondonPRL2013,LiuNanoscale2014}.

In this paper we present an analytical formula and a comprehensive
understanding on how to optimize the DNP by an optically pumped NV center
near the GSLAC at ambient temperature. Our results provide a parameter-free
explanation and a clear physics picture for the experimentally observed
magnetic field dependence of $^{13}$C nuclear polarization \cite%
{WangNatCommun2013}. More importantly, we demonstrate the possibility to
greatly enhance this magnetic field dependence and strongly polarize weakly
coupled $^{13}$C nuclei (with HFI down to $\sim 1$ kHz) via direct, resonant
spin transfer under suitable conditions. First, we introduce our model and
an intuitive picture for manipulating the DNP with an optically pumped NV
center. Then we present our theory and analytical formula for the DNP of a
single nuclear spin, which reveals the possibility of strongly polarizing
weakly coupled nuclear spins under weak optical pumping and fine-tuned
magnetic field. Finally, we demonstrate this possibility for multiple
nuclear spins by providing a multi-spin DNP theory and its numerical
solution.

\section*{Results}

\begin{figure}[tbp]
\includegraphics[width=\columnwidth]{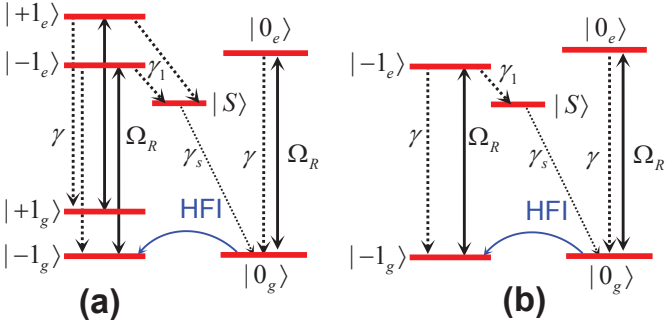}
\caption{(a) NV states at ambient temperature responsible for DNP under
optical pumping. (b) Reduced five-level model for DNP near the ground state
anticrossing.}
\label{G_ENERGYLEVEL}
\end{figure}

\subsection*{Model and intuitive physics picture}

First we briefly introduce our theoretical model, consisting of a negatively
charged NV center coupled to many surrounding nuclear spins at ambient
temperature. The NV center has a ground state triplet $|\pm 1_{g}\rangle $
and $|0_{g}\rangle $ separated by zero-field splitting $D_{\mathrm{gs}}=2.87$
$\mathrm{GHz}$ and an excited state triplet $|\pm 1_{e}\rangle $ and $%
|0_{e}\rangle $ separated by zero-field splitting $D_{\mathrm{es}}=1.41\
\mathrm{GHz}$ \cite{FuchsPRL2008}. In a magnetic field $B$ along the N-V
axis ($z$ axis), the electron Zeeman splitting $\gamma_{e}B$ with $\gamma
_{e}=28.025\ $\textrm{GHz/T} cancels the ground state zero-field splitting
at a critical magnetic field $D_{\mathrm{gs}}/\gamma_{e}\approx 102\
\mathrm{mT}$, leading to GSLAC between $|0_{g}\rangle $ and $|-1_{g}\rangle $%
. The GSLAC reduces the energy mismatch for the NV-nuclei flip-flop dynamics
and enables NV-induced DNP through their HFI $\hat{H}_{\mathrm{HF}}=\sum_{i}%
\hat{\mathbf{F}}_{i}\cdot \hat{\mathbf{I}}_{i}$, where $\hat{\mathbf{F}}%
_{i}\equiv \hat{\mathbf{S}}_{g}\cdot \mathbf{A}_{g,i}+\hat{\mathbf{S}}%
_{e}\cdot \mathbf{A}_{e,i}$ is the electron Knight field coupled to the $i$%
th nucleus, and $\hat{\mathbf{S}}_{g}$ ($\hat{\mathbf{S}}_{e}$) is the NV
ground (excited) state spin. The HFI tensors $\mathbf{A}_{g,i}$ and $\mathbf{%
A}_{e,i}$ are well established for the on-site nitrogen nucleus \cite%
{DohertyPhysRep2013} and the nearest $^{13}\mathrm{C}$ nuclear spins in the
first shell of the vacancy \cite%
{LoubserRPP1978,GaliPRB2008,GaliPRB2009a,FeltonPRB2009}. For other $^{13}$C
nuclei, especially those $>6.3$ \textrm{\AA } away from the NV center, very
little information is available \cite%
{GaliPRB2008,FeltonPRB2009,SmeltzerNJP2011,DreauPRB2012} and dipolar HFI is
usually assumed \cite%
{MazePRB2008,ZhaoPRB2012,CappellaroPRA2012,LondonPRL2013,LiuNanoscale2014}.

Next we provide an intuitive physics picture for engineering the DNP by
controlling the NV center near the GSLAC. For brevity we focus on one
nuclear spin-$I$ (e.g., $I=1/2$ for the $^{13}$C or $^{15}$N nucleus and $%
I=1 $ for the $^{14}$N nucleus) and drop the nuclear spin index $i$. To
describe the NV-nucleus flip-flop, we decompose the HFI into the
longitudinal part $\hat{F}_{z}\hat{I}_{z}$ that conserves the nuclear spin
and the transverse part $(\hat{F}_{+}\hat{I}_{-}+h.c.)/2$ that induces
NV-nuclear spin flip-flop dynamics. As we always work near the GSLAC, the
nuclear spin flip induced by the off-resonant excited state HFI is very
small (to be discussed shortly). Since $|-1_{g}\rangle $ is nearly
degenerate with the NV steady state $|0_{g}\rangle $ under optical pumping,
the NV-nucleus flip-flop is dominated by the $\hat{\sigma}_{-1_{g},0_{g}}%
\hat{I}_{+}$ process $|0_{g}\rangle \otimes |m\rangle \longrightarrow
|-1_{g}\rangle \otimes |m+1\rangle $ and the $\hat{\sigma}_{-1_{g},0_{g}}%
\hat{I}_{-}$ process $|0_{g}\rangle \otimes |m+1\rangle \longrightarrow
|-1_{g}\rangle \otimes |m\rangle $, where $\{|m\rangle \}$ are nuclear spin
Zeeman eigenstates. The corresponding energy mismatches of these two
processes are
\begin{align*}
\Delta _{m+1\leftarrow m}& \equiv \Delta +\gamma _{N}B-(m+1)A_{g,zz}, \\
\Delta _{m\leftarrow m+1}& \equiv \Delta -\gamma _{N}B-mA_{g,zz},
\end{align*}%
respectively, where the first term $\Delta \equiv D_{\mathrm{gs}}-\gamma
_{e}B$ is the $|-1_{g}\rangle $-$|0_{g}\rangle $ energy separation, the
second term $\pm \gamma _{N}B$ comes from the nuclear spin Zeeman term,
while the last term comes from the longitudinal HFI $\hat{F}_{z}\hat{I}%
_{z}\approx -(A_{g,zz}\hat{\sigma}_{-1_{g},-1_{g}}+A_{e,zz}\hat{\sigma}%
_{-1_{e},-1_{e}})\hat{I}_{z}$. The energy mismatch difference
\begin{equation}
\Delta _{N}\equiv \Delta _{m+1\leftarrow m}-\Delta _{m\leftarrow
m+1}=2\gamma _{N}B-A_{g,zz}  \label{DELTAN}
\end{equation}%
provide the first essential ingredient for engineering the DNP by
selectively driving one process into resonance.

The second essential ingredient for engineering the dissipative DNP is
optical pumping, which connects the NV center to the strongly dissipative
bath of vacuum electromagnetic fluctuation and makes the NV center itself a
dissipative, non-equilibrium bath: it quickly establishes the unique steady
state $|0_{g}\rangle $ whatever the initial state is (commonly known as
optical initialization). When optical pumping is not too weak, such that the
NV dissipation (more precisely the NV optical initialization) is much faster
than nuclear spin dissipation (more precisely the DNP process), the NV
center can be regarded as always in its steady state $|0_{g}\rangle $, i.e.,
it becomes a\textit{\ }Markovian bath and induces DNP by driving the two
processes $\hat{\sigma}_{-1_{g},0_{g}}\hat{I}_{\pm }$ \cite%
{WangNatCommun2013}. The resonance linewidth of both processes are
determined by the optically induced NV ground state level broadening.
Therefore, under sufficiently weak (but not too weak) optical pumping such
that the linewidth of each resonance is much narrower than the distance $%
|\Delta _{N}|$ between the two resonances, we can fine tune the magnetic
field to drive $\hat{\sigma}_{-1_{g},0_{g}}\hat{I}_{+}$ ($\hat{\sigma}%
_{-1_{g},0_{g}}\hat{I}_{-}$) into resonance, while keeping $\hat{\sigma}%
_{-1_{g},0_{g}}\hat{I}_{-}$ ($\hat{\sigma}_{-1_{g},0_{g}}\hat{I}_{+}$)
off-resonance to achieve strong positive (negative) nuclear polarization.
When the magnetic field is swept from $\hat{\sigma}_{-1_{g},0_{g}}\hat{I}%
_{+} $ resonance to $\hat{\sigma}_{-1_{g},0_{g}}\hat{I}_{-}$ resonance, the $%
^{13} $C nuclear polarization changes from positive values to negative
values over a magnetic field range $\sim |\Delta _{N}|/\gamma_{e}$. For
first-shell $^{13}$C nuclei, $|\Delta _{N}|$ is dominated by $%
|A_{g,zz}|\approx 130$ MHz. This gives a simple explanation to the
experimentally observed polarization sign reversal over a few mT \cite%
{WangNatCommun2013}. For weakly coupled $^{13}$C nuclei, $|\Delta
_{N}|\approx 2|\gamma _{N}B|\approx 2.2$ MHz is greatly reduced,
corresponding to a much more sensitive dependence of the $^{13}$C
polarization on the magnetic field over a range $\sim 0.1$ mT. Below we
demonstrate this intuitive idea and provide a complete picture for the DNP
of a single nuclear spin-$I$ before going into the more complicated case of
multi-nuclei DNP.

\subsection*{DNP theory of single nuclear spin}

Under optical pumping from the NV ground orbital $|g\rangle $ to the excited
orbital $|e\rangle $, seven NV states are relevant, including the ground
state triplet, excited state triplet, and a metastable singlet state $%
|S\rangle $ [see Fig. \ref{G_ENERGYLEVEL}(a)]. The NV Hamiltonian consists
of the orbital part $\omega _{0}\hat{\sigma}_{e,e}$, the optical pumping
term $\hat{H}_{c}(t)=(\Omega _{R}/2)e^{-i\omega t}\hat{\sigma}_{e,g}+h.c.$,
the ground state triplet part $\hat{H}_{\mathrm{gs}}\equiv D_{\mathrm{gs}}%
\hat{S}_{g,z}^{2}+\gamma _{e}B\hat{S}_{g,z}$, and the excited state triplet
part $\hat{H}_{\mathrm{es}}\equiv D_{\mathrm{es}}\hat{S}_{e,z}^{2}+\gamma
_{e}B\hat{S}_{e,z}$, where $\hat{\sigma}_{i,j}\equiv |i\rangle \langle j|$
is the transition operator. In the rotating frame of the optical pumping,
the NV-nucleus coupled system obeys
\begin{equation}
\dot{\hat{\rho}}(t)=\mathcal{L}_{\mathrm{NV}}\hat{\rho}(t)-i[\hat{H}_{N}+%
\hat{\mathbf{F}}\cdot \hat{\mathbf{I}},\hat{\rho}(t)],  \label{LINDBLAD}
\end{equation}%
where $\hat{H}_{N}=\gamma _{N}B\hat{I}_{z}$ is the nuclear spin Zeeman
Hamiltonian and $\mathcal{L}_{\mathrm{NV}}(\cdot )\equiv -i[\hat{H}_{\mathrm{%
NV}},(\cdot )]+\sum_{\alpha }\mathcal{D}[\hat{L}_{\alpha }](\cdot )$ is the
evolution superoperator of the NV center in the absence of the nuclear spin,
with $\hat{H}_{\mathrm{NV}}$ the NV Hamiltonian in the rotating frame and
the last term accounting for various NV dissipation channels [shown in Fig. %
\ref{G_ENERGYLEVEL}(a)] in the Lindblad form $\mathcal{D}[\hat{L}]\hat{\rho}%
\equiv (\hat{L}\hat{\rho}\hat{L}^{\dagger }-\{\hat{L}^{\dagger }\hat{L},\hat{%
\rho}\}/2)$, as well as the excited orbital pure dephasing at a rate $\Gamma
_{e}$ and the ground state spin dephasing at a rate $\gamma _{\varphi
}=1/T_{2,\mathrm{NV}}$ that models the finite ground state coherence time $%
T_{2,\mathrm{NV}}$. We use the experimentally measured dissipation rates at
room temperature:$\ \Gamma_{e}=10^{4}\ \mathrm{GHz}$ \cite{AbtewPRL2011}, $%
\gamma =13$ $\mathrm{MHz}$ \cite{MansonPRB2006}, $\gamma _{1}\approx 13.3\
\mathrm{MHz}$ \cite{RobledoNJP2011}, and $\gamma _{s}=0.56\ \mathrm{MHz}$
\cite{MaPRB2010,ChoiPRB2012,AcostaPRB2010,DelaneyNanoLett2010}. Here we have
neglected the very small leakage via intersystem crossing from $%
|0_{e}\rangle $ to $|\pm 1_{g}\rangle $, consistent with the experimentally
reported \cite{JiangScience2009,ToganNature2010} high optical initialization
probability $\sim 96\%$ into $|0_{g}\rangle $. Indeed, we have verified that
the nuclear polarization is not so sensitive to the optical initialization
probability, e.g., upon including appreciable leakage such that the optical
initialization probability drops to 80\%, the steady-state $^{13}$C nuclear
polarization drops by $\sim 20\%$.

The DNP can be directly obtained by solving Eq. (\ref{LINDBLAD})
numerically. However, this approach does not provide a clear physics picture
for the underlying DNP mechanism and is not suitable for searching for
optimal experimental parameters to maximize the DNP effect. Further, for the
DNP of multiple nuclear spins (to be discussed shortly), this approach
quickly becomes infeasible with increasing number $N$ of the nuclei, because
the dimension of the Liouville space grows exponentially as $\dim(\hat{\rho }%
)\approx(2I+1)^{2N}M^{2}$, where $M=7$ is the number of relevant energy
levels of the NV center, e.g., the DNP of $N=5$ nuclear spin-1/2's already
involves $\dim(\hat{\rho})\approx50000$ and hence is computationally
intensive.

Our work is based on a recently developed microscopic theory \cite%
{YangPRB2013,YangPRB2012,WangArxiv2015}, applicable as long as the NV
dissipation (optical initialization in the present case) is much faster than
the nuclear spin dissipation (DNP in the present case). The populations $%
\{p_{m}\}$ of a single nuclear spin-$I$ on its Zeeman sublevels $|m\rangle $
($m=-I,-I+1,\cdots ,I$) obey the rate equations%
\begin{equation}
\dot{p}_{m}=\sum_{n=m\pm 1}W_{m\leftarrow n}p_{n}-(\sum_{n=m\pm
1}W_{n\leftarrow m})p_{m}.  \label{RATE_EQ}
\end{equation}%
Up to second order of the \textit{tranvserse} HFI (longitudinal HFI treated
\emph{exactly}), the transition rate from $|m\rangle $ to $|m\pm 1\rangle $
is \cite{WangArxiv2015}
\begin{equation}
W_{m\pm 1\leftarrow m}=\frac{\xi _{m}^{\pm }}{2}\mathrm{Re}\int_{0}^{+\infty
}dt\ e^{\mp i\gamma _{N}Bt}\mathrm{Tr}_{\mathrm{NV}}\hat{F}_{\mp }^{\dagger
}e^{\mathcal{L}_{m\pm 1,m}t}\hat{F}_{\mp }\hat{P}_{m},  \label{WPM}
\end{equation}%
where $\xi _{m}^{\pm }\equiv \langle m|\hat{I}_{\mp }\hat{I}_{\pm }|m\rangle
$, the Liouville superoperator $\mathcal{L}_{n,m}(\bullet )\equiv \mathcal{L}%
_{\mathrm{NV}}(\bullet )-i[n\hat{F}_{z}(\bullet )-(\bullet )m\hat{F}_{z}]$,
and $\hat{P}_{m}$ is the NV steady state in the rotating frame conditioned
on the nuclear spin state being $|m\rangle $, i.e., $\mathcal{L}_{m,m}\hat{P}%
_{m}=0$ and $\mathrm{Tr}_{\mathrm{NV}}\hat{P}_{m}=1$. The nuclear spin
depolarization by other mechanisms such as spin-lattice relaxation can be
described by a phenomenological depolarization rate $\gamma _{\mathrm{dep}}$
and incorporated by replacing $W_{m\pm 1\leftarrow m}$ with $W_{m\pm
1\leftarrow m}+\xi _{m}^{\pm }\gamma _{\mathrm{dep}}$.

Equation (\ref{WPM}) can be understood as a generalized
fluctuation-dissipation relation \cite{ClerkRMP2010}. Since the nuclear spin
transition from $|m\rangle$ to $|m+1\rangle$ ($|m-1\rangle$) is induced by
the HFI term $\hat{F}_{-}\hat{I}_{+}$ ($\hat{F}_{+}\hat{I}_{-}$) and
involves the transfer of energy $\gamma_{N}B$ from the NV center (nuclear
spin) to the nuclear spin (NV center), the corresponding transition rate is
proportional to the fluctuation of the Knight field $\hat{F}_{-}$ ($\hat{F}%
_{+}$) at frequency $-\gamma_{N}B$ ($+\gamma_{N}B$) in the NV steady state $%
\hat{P}_{m}$. Without the longitudinal HFI $\hat{F}_{z}\hat{I}_{z}$, we have
$\mathcal{L}_{n,m}=\mathcal{L}_{\mathrm{NV}}$, so $\hat{P}_{m}=\hat{P}$ is
the steady NV state in the absence of the nuclei, and Eq. (\ref{WPM})
reduces to the conventional Born-Markovian approximation. As discussed above
and further elaborated below, the longitudinal HFI $\hat{F}_{z}\hat{I}_{z}$
could significantly modify the energy cost of the nuclear spin flip and
hence the steady-state nuclear polarization.

In general, the nuclear spin transition rates $W_{m\pm1\leftarrow m}$ in Eq.
(\ref{WPM}) can be evaluated numerically (see Methods). However, the
physical picture and global optimization of the experimental parameters will
be greatly facilitated by explicit analytical expressions, which we present
below.

\subsection*{Explicit analytical expressions}

Since we always consider $B\sim 102$ mT near $|0_{g}\rangle $-$%
|-1_{g}\rangle $ GSLAC, the dominant NV-nucleus flip-flop involves the
near-resonant process $|0_{g}\rangle \leftrightarrow |-1_{g}\rangle $. This
allows us to neglect $|+1_{g}\rangle $ and $|+1_{e}\rangle $, describe the
NV center by a five-level model [Fig. \ref{G_ENERGYLEVEL}(b)], and derive
explicit analytical expressions for the nuclear spin transition rates $%
W_{m\pm 1\leftarrow m}$ (see Methods). For general parameters, the
expressions are complicated due to the presence of various quantum coherence
effects associated with the optical pumping induced $|0_{g}\rangle $-$%
|0_{e}\rangle $ and $|-1_{g}\rangle $-$|-1_{e}\rangle $ oscillation.
However, if we focus on weak optical pumping far from saturation, i.e.,
optical pumping rate $R\equiv \Omega _{R}^{2}/\Gamma_{e}$ $\ll $ linewidth $%
\gamma +\gamma _{1}/2\approx 26.3$ MHz of $|0_{e}\rangle \leftrightarrow
|-1_{e}\rangle $ transition, then the quantum coherence effects are
negligible and $W_{m\pm 1\leftarrow m}$ assume the appealing form of a Fermi
golden rule%
\begin{equation}
W_{m\pm 1\leftarrow m}=P_{g}\xi _{m}^{\pm }2\pi \left\vert \frac{A_{g,+\mp }%
}{2\sqrt{2}}\right\vert ^{2}\delta ^{(\Gamma )}(\Delta _{m\pm 1\leftarrow
m}),  \label{WPM_5LEVEL}
\end{equation}%
where $P_{g}=(R+\gamma )/(2R+\gamma )$ is the steady-state populations on $%
|0_{g}\rangle $, $\delta ^{(\gamma )}(x)\equiv (\gamma /\pi )/(x^{2}+\gamma
^{2})$ is the Lorentzian shape function, $\Gamma \equiv \gamma _{\varphi }+R$
is the resonance linewidth, and $A_{g,+\mp }\equiv \mathbf{e}_{+}\cdot
\mathbf{A}_{g}\cdot \mathbf{e}_{\mp }$ ($\mathbf{e}_{\pm }\equiv \mathbf{e}%
_{x}\pm i\mathbf{e}_{y}$) quantifies the coefficient of the term $\hat{\sigma%
}_{-1_{g},0_{g}}\hat{I}_{\pm }$ in the transverse HFI. A key feature of Eq. (%
\ref{WPM_5LEVEL}) is that the resonance linewidth of both nuclear spin-flip
processes is the sum of $|0_{g}\rangle $-$|-1_{g}\rangle $ decoherence rate $%
\gamma _{\varphi }=1/T_{2,\mathrm{NV}}$ ($\sim$ 1 kHz $\ll $ typically $R$)
and optical pumping induced NV level broadening (which equals the optical
pumping rate $R$). This result provides a clear understanding of the laser
induced NV ground state level broadening previously observed in the
cross-relaxation between the NV center and nearby nitrogen impurities \cite%
{HansonPRL2006,GaebelNature2006}. Further, it indicates that under
sufficiently weak pumping, the opposite processes $W_{m+1\leftarrow m}$ and $%
W_{m\leftarrow m+1}$ can be selectively driven into resonance to achieve
strong nuclear spin polarization. Equation (\ref{WPM_5LEVEL}) also reveals
that the DNP depends strongly on the form of the HFI tensor $\mathbf{A}_{g}$%
, e.g., for the $^{14}$N or $^{15}$N nucleus with isotropic transverse HFI $%
A_{g,xx}=A_{g,yy}$ and $A_{g,xy}=0$, we have $A_{g,++}=0$ and $%
A_{g,+-}=2A_{g,xx}$ and hence complete positive polarization.

Equation (\ref{WPM_5LEVEL}) can be applied to a general nuclear spin-$I$,
but hereafter we focus on a $^{13}$C nuclear spin-1/2 and define $%
W_{+}\equiv W_{\uparrow \leftarrow \downarrow }$ and $W_{-}\equiv
W_{\downarrow \leftarrow \uparrow }$, given by Eq. (\ref{WPM_5LEVEL}) as%
\begin{equation}
W_{\pm }=P_{g}2\pi \left\vert \frac{A_{g,+\mp }}{2\sqrt{2}}\right\vert
^{2}\delta ^{(\Gamma )}(\Delta \pm \Delta _{N}/2).  \label{WPM_5LEVEL_SPIN12}
\end{equation}%
The evolution of the nuclear polarization $p\equiv 2\langle \hat{I}%
_{z}\rangle $ follows from $\dot{p}(t)=-W[p(t)-p_{\mathrm{ss}}]$ as $p(t)=p_{%
\mathrm{ss}}+[p(0)-p_{\mathrm{ss}}]e^{-Wt}$, so the DNP of a $^{13}$C
nuclear spin is completely determined by two key quantities: the
steady-state polarization $p_{\mathrm{ss}}\equiv (W_{+}-W_{-})/(W_{+}+W_{-})$
and the rate $W\equiv W_{+}+W_{-}$ of DNP. Up to now, we have neglected
other nuclear spin relaxation mechanisms, such as spin-lattice relaxation
and transverse HFI with NV excited states. The former occurs on the time
scale ranging from a few seconds to tens of minutes \cite%
{MaurerScience2012,FischerPRL2013}. The latter occurs on the time scale $%
1/W_{\mathrm{es}}$, where $W_{\mathrm{es}}$ can be estimated from the Fermi
golden rule as $2\pi (A_{e}/2)^{2}\delta ^{(\gamma +\gamma _{1}/2)}(D_{%
\mathrm{es}})P_{e}$, e.g., for a $^{13}$C located at $|\mathbf{R}|>3\
\mathrm{\mathring{A}}$ coupled to the NV excited states via dipolar HFI, we
have $1/W_{\mathrm{es}}>1$ s for $R=0.2\ \mathrm{MHz}$. These effects are
described by nuclear depolarization $W_{\pm }\rightarrow W_{\pm }+\gamma _{%
\mathrm{dep}}$, which increases $W$ by $2\gamma _{\mathrm{dep}}$ and
decreases $p_{\mathrm{ss}}$ by a factor $1+2\gamma _{\mathrm{dep}}/W$, i.e.,
nuclear depolarization is negligible when $W\gg 2\gamma _{\mathrm{dep}}$.

Finally we emphasize that our rate equation theory and hence Eq. (\ref%
{WPM_5LEVEL_SPIN12}) are accurate when the DNP time $1/W$ from Eq. (\ref%
{WPM_5LEVEL_SPIN12}) $\gg $ NV optical initialization time $\tau _{c}\approx
([R\gamma _{1}/(\gamma _{1}+\gamma )]^{-1}+\gamma _{s}^{-1}$. When the
optical pumping is so weak that the calculated DNP time from Eq. (\ref%
{WPM_5LEVEL_SPIN12}) drops below $\tau _{c}$, the NV center becomes a
non-Markovian bath and the true DNP time would be lower bounded by $\sim
\tau _{c}$, as we discuss below.

\subsection*{DNP of strongly coupled $^{13}\mathrm{C}$ nucleus}

\begin{figure}[tbp]
\includegraphics[width=\columnwidth]{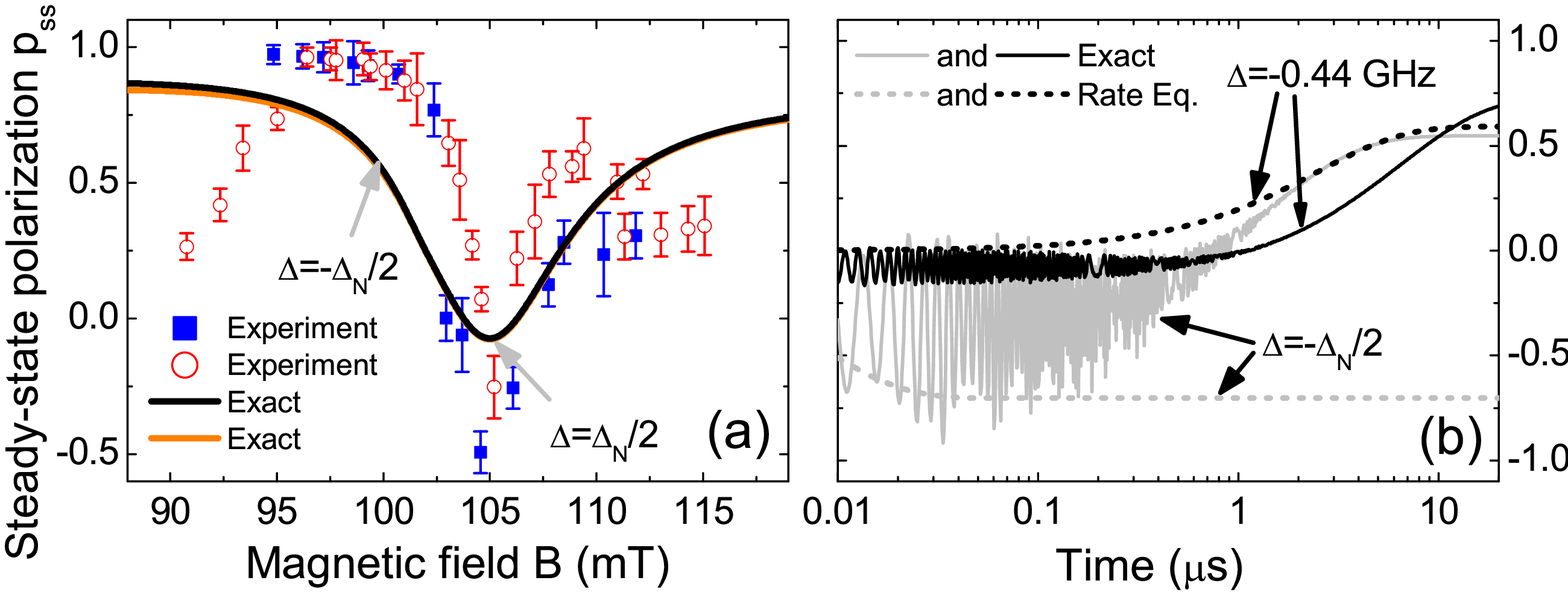}
\caption{ (a) $p_{\mathrm{ss}}$ of a first-shell $^{13}$C nucleus: exact
solution to Eq. (\protect\ref{LINDBLAD}) with (orange solid line) or without
(black solid line) excited state HFI, compared with the experimental data
\protect\cite{WangNatCommun2013} (filled squares and empty circles) from two
different analysis methods. (b) Real-time evolution of nuclear polarization $%
p(t)$ on $W_{+}$ resonance ($\Delta =-\Delta _{N}/2$) or far from the
resonance ($\Delta =-0.44$ GHz): exact solution to Eq. (\protect\ref%
{LINDBLAD}) vs. our rate equation theory [Eqs. (\protect\ref{RATE_EQ}) and (%
\protect\ref{WPM})].}
\label{G_COMPARE_EXP}
\end{figure}

To begin with, we consider the DNP of a strongly coupled $^{13}$C nucleus in
the first shell of the NV center under optical pumping near the $%
|0_{g}\rangle $-$|-1_{g}\rangle $ GSLAC. This configuration has been studied
experimentally in ensembles of NV centers \cite{WangNatCommun2013}, which
reveals an interesting polarization sign reversal over a narrow range of
magnetic fields. This is attributed to the anisotropic HFI and could allow
magnetic control of the $^{13}$C nuclear polarization, but a quantitative
understanding remains absent. Here we provide a clear physics picture and
parameter-free explanation for this sign reversal using the experimentally
measured ground state HFI tensor \cite%
{LoubserRPP1978,FeltonPRB2009,GaliPRB2009a,WangNatCommun2013} $%
A_{g,xx}=198.6\ \mathrm{MHz}$, $A_{g,yy}=123.0$ $\mathrm{MHz}$, $%
A_{g,zz}=129.0$ $\mathrm{MHz}$, $A_{g,xz}=A_{g,zx}=-21.5$ $\mathrm{MHz}$,
and excited state HFI tensor \cite{GaliPRB2009a} $A_{e,xx}=103.2$ $\mathrm{%
MHz}$, $A_{e,yy}=56.7\ \mathrm{MHz}$, $A_{e,zz}=79.5\ \mathrm{MHz}$, $%
A_{e,xz}=A_{e,zx}=-32.6\ \mathrm{MHz}$. To focus on the intrinsic behavior
of DNP induced by the ground state HFI, we set $\gamma _{\mathrm{dep}}=0$.
We have verified that due to the strong HFI induced NV-$^{13}$C mixing, the
magnetic field dependence of the $^{13}$C nuclear polarization depends
weakly on the optical pumping up to $R<4$ MHz. In our numerical calculation,
we take $R=0.4$ MHz.

First we discuss the exact solution to the Lindblad master equation Eq. (\ref%
{LINDBLAD}) without the excited state HFI. In Fig. \ref{G_COMPARE_EXP}(a),
the calculated steady-state polarization $p_{\mathrm{ss}}$ (black solid
lines) agrees reasonably with the experimental data \cite{WangNatCommun2013}%
. According to our analytical expression Eq. (\ref{WPM_5LEVEL_SPIN12}), the
negative dip at $B\approx 105$ mT and positive shoulder at $B\sim 100$ mT
correspond, respectively, to $W_{-}$ resonance at $\Delta =\Delta _{N}/2$
and $W_{+}$ resonance at $\Delta =-\Delta _{N}/2$. The strong transverse HFI
leads to rapid DNP and significant non-Markovian oscillation in the
real-time evolution of the nuclear polarization $p(t)$ [solid lines in Fig. %
\ref{G_COMPARE_EXP}(b)], beyond the description of our rate equation theory.
Going successively closer to GSLAC, the DNP time shortens [Fig. \ref%
{G_COMPARE_EXP}(b)] and finally saturates at $\sim 1.6$ $\mathrm{\mu s}$,
lower bounded by the NV optical initialization time $\tau _{c}\approx 1.1\
\mathrm{\mu s}$. The small difference between the black and orange solid
lines in Fig. \ref{G_COMPARE_EXP}(a) confirms that the far off-resonant
excited state HFI has a negligible effect. Indeed, the exact numerical
solution to Eq. (\ref{LINDBLAD}) including the excited state HFI shows that
over the whole range of magnetic field under consideration, the excited
state HFI alone polarizes the nuclear spin towards $p_{\mathrm{ss}}\sim 0.5$
with a DNP time $\sim 30$-$40\ \mathrm{\mu s}$ (much slower than the ground
state HFI induced DNP), consistent with the estimation based on the Fermi
golden rule.
\begin{figure}[tbp]
\includegraphics[width=\columnwidth]{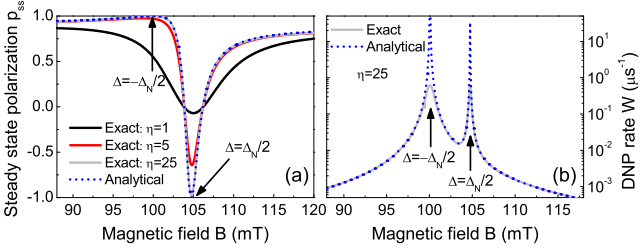}
\caption{ (a) Steady-state polarization $p_{\mathrm{ss}}$ of a first-shell $%
^{13}$C nucleus: exact numerical solution to Eq. (\protect\ref{LINDBLAD})
(solid lines) vs. analytical formula Eq. (\protect\ref{WPM_5LEVEL_SPIN12})
(dotted line). To illustrate the convergence of our theory, we scale down
the transverse HFI ($A_{g,xx}$ and $A_{g,yy}$) by a factor $\protect\eta =1$
(black line), $5$ (red line), and $25$ (gray line), with $%
A_{g,xz}=A_{g,zx}=0 $ for $\protect\eta =5$ and $25$ to avoid spurious
behaviors. (b) DNP rate for $\protect\eta =25$:\ exact solution to Eq. (%
\protect\ref{LINDBLAD}) (gray solid line) vs. analytical formula Eq. (%
\protect\ref{WPM_5LEVEL_SPIN12}) (blue dotted line).}
\label{G_EXACT_RATEEQ}
\end{figure}

Next we demonstrate that our analytical expression Eq. (\ref%
{WPM_5LEVEL_SPIN12}) is accurate as long as the Markovian condition $W\ll
1/\tau _{c}$ is satisfied, e.g., for magnetic field away from GSLAC or for $%
^{13}$C nuclei with weaker transverse HFI. For this purpose, we manually
scale down $A_{g,xx}$ and $A_{g,yy}$ by a factor $\eta $ to decrease the DNP
rate, but keep the longitudinal component $A_{g,zz}$ invariant. In Fig. \ref%
{G_EXACT_RATEEQ}(a), the analytical nuclear polarization $p_{\mathrm{ss}}$
agrees well with the exact solution to Eq. (\ref{LINDBLAD}) for $\eta
\gtrsim 5$. In Fig. \ref{G_EXACT_RATEEQ}(b), when approaching $W_{\pm }$
resonances, the analytical DNP rate from Eq. (\ref{WPM_5LEVEL_SPIN12})
sharply peaks and significantly exceeds $1/\tau _{c}\approx (1.1\ \mathrm{%
\mu s})^{-1}$, while the exact solution saturates to $\sim $ $(1.6\ \mathrm{%
\mu s})^{-1}$. Away from $W_{\pm }$ resonances, our analytical DNP rate $\ll
$ $1/\tau _{c}$ and hence agrees well with the exact solution.

In the above, for a first-shell $^{13}$C nuclear spin, the separation $%
\Delta _{N}=2\gamma _{N}B-A_{g,zz}\approx -131$ MHz is dominated by the
strong HFI term. Consequently, the magnetic fields for the two resonances at
$\Delta =\pm \Delta _{N}/2$ are separated by $\Delta _{N}/(\gamma
_{e}B)\approx 5$ mT, which determines the magnetic field sensitivity of the
polarization sign reversal in Fig. \ref{G_EXACT_RATEEQ}(a). This suggests
that for weakly coupled $^{13}$C nuclear spins with much smaller $\Delta
_{N}\approx -2$ MHz, the magnetic field sensitivity can be greatly enhanced.

\subsection*{DNP of weak coupled $^{13}\mathrm{C}$ nucleus}

Now based on the analytical expression Eq. (\ref{WPM_5LEVEL_SPIN12}), we
provide a complete picture for the dependence of the DNP rate $W$ and
steady-state polarization $p_{\mathrm{ss}}$ on the resonance linewidth $%
\Gamma \approx R$, the $|0_{g}\rangle $-$|-1_{g}\rangle $ separation $\Delta
=D_{\mathrm{gs}}-\gamma _{e}B$, and the hyperfine tensor $\mathbf{A}_{g}$.
It reveals the possibility to strongly polarize a weakly coupled, distant $%
^{13}$C nucleus.

\begin{figure}[tbp]
\includegraphics[width=\columnwidth]{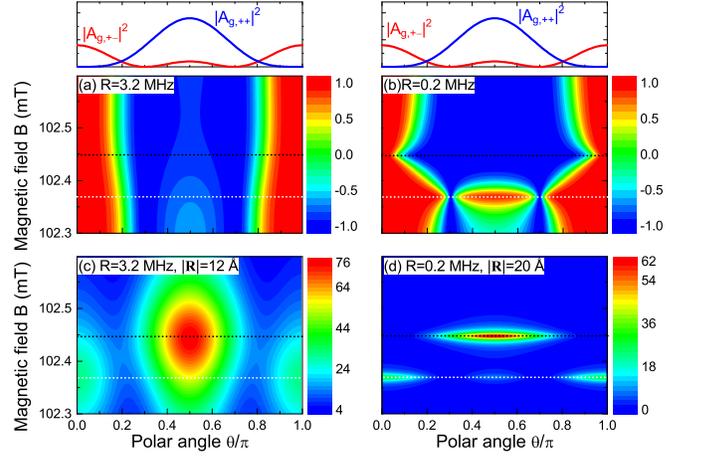}
\caption{Steady state nuclear polarization $p_{\mathrm{ss}}$ [(a) and (b)]
and DNP rate $W$ in units of Hz [(c) and (d)] by dipolar HFI with the NV
center, calculated from the analytical formula Eq. (\protect\ref%
{WPM_5LEVEL_SPIN12}) with optical pumping rates $R=3.2$ MHz [(a) and (c)]
and $0.2$ MHz [(b) and (d)]. The distance of the $^{13}$C nucleus from the
NV center is $|\mathbf{R}|=12$ \AA\ in (c) and $|\mathbf{R}|=20$ \AA\ in
(d). The white (black) dotted line indicates the critical magnetic field for
$W_{+}$ ($W_{-}$) resonance.}
\label{G_WEAK_CONTOUR}
\end{figure}

The simplest case is strong optical pumping with the resonance linewidth $%
\Gamma >|\Delta _{N}|$, i.e., the $W_{\pm }$ resonance peaks at $\Delta
=-\Delta _{N}/2$ and $\Delta =\Delta _{N}/2$, respectively, are not
spectrally resolved: $W_{\pm }\approx \pi (|A_{g,+\mp }|^{2}/4)\delta
^{(\Gamma )}(\Delta )$. In this case $p_{\mathrm{ss}}$ is uniquely
determined by the HFI tensor $\mathbf{A}_{g}$, but $W$ still depends on the $%
|0_{g}\rangle $-$|-1_{g}\rangle $ separation $\Delta $ and becomes
suppressed when $|\Delta |\gg \Gamma $. For dipolar HFI,
\begin{align*}
A_{g,zz}& =-A_{g,+-}=A_{\mathrm{dp}}(|\mathbf{R}|)(1+c_{\theta }), \\
A_{g,++}& =A_{\mathrm{dp}}(|\mathbf{R}|)(1-c_{\theta })e^{2i\varphi },
\end{align*}%
where $A_{\mathrm{dp}}(|\mathbf{R}|)\equiv -\mu _{0}\gamma _{e}\gamma
_{N}/(4\pi |\mathbf{R}|^{3})\approx 20\ \mathrm{MHz}\ $\textrm{\AA }$^{3}/|%
\mathbf{R}|^{3}$, $\gamma _{N}=-10.705$ \textrm{MHz/T} is the $^{13}$C
nuclear gyromagnetic ratio, $c_{\theta }\equiv 3\cos ^{2}\theta -2$, and $%
\theta $ $(\varphi )$ is the polar (azimuth) angle of the $^{13}$C location $%
\mathbf{R}$. Thus the polarization
\begin{equation}
p_{\mathrm{ss}}\approx \frac{2}{c_{\theta }+1/c_{\theta }}
\label{PSS_DIPOLAR}
\end{equation}%
is uniquely determined by the polar angle $\theta $ and is independent of $|%
\mathbf{R}|$ as long as the DNP rate $W\gg 2\gamma _{\mathrm{dep}}$ or
equivalently%
\begin{equation}
A_{\mathrm{dp}}^{2}(|\mathbf{R}|)\gg 4\Gamma \gamma _{\mathrm{dep}},
\label{STRONG_HFI}
\end{equation}%
e.g., for $\Gamma =0.2\ \mathrm{MHz}$ and $\gamma _{\mathrm{dep}}=1$ s$^{-1}$%
, this condition amounts to $A_{\mathrm{dp}}>1$ kHz and is satisfied by
nuclei with $|\mathbf{R}|<27$ \textrm{\AA }. An example for strong optical
pumping is shown in Figs. \ref{G_WEAK_CONTOUR}(a) and \ref{G_WEAK_CONTOUR}%
(c). In Fig. \ref{G_WEAK_CONTOUR}(c), the $W_{+}$ resonance peak (marked by
white dotted lines) and the $W_{-}$ resonance peak (marked by black dotted
lines) are significantly broadened along the magnetic field axis and are not
clearly resolved. Consequently, the nuclear polarization in Fig. \ref%
{G_WEAK_CONTOUR}(a) follows the HFI coefficients $|A_{g,+\mp }|^{2}\propto
(1\pm c_{\theta })^{2}$ [plotted on top of Figs. \ref{G_WEAK_CONTOUR}(a) and %
\ref{G_WEAK_CONTOUR}(b)] and is well described by Eq. (\ref{PSS_DIPOLAR}).

The most interesting case is weak optical pumping with the resonance
linewidth $\Gamma \ll |\Delta _{N}|$. In this case, the $W_{\pm }$ resonance
peaks are spectrally resolved and can be selectively driven into resonance
to achieve strong nuclear polarization $p_{\mathrm{ss}}=\pm \Delta
_{N}^{2}/(\Delta _{N}^{2}+2\Gamma ^{2})\approx \pm 100\%$ by choosing $%
\Delta =\mp \Delta _{N}/2$. The DNP rate $W\propto 1/\Gamma $ is dominated
by the resonant process and can be increased up to the saturation value $%
\sim 1/\tau _{c}$ by decreasing the linewidth $\Gamma $. An example is given
in Figs. \ref{G_WEAK_CONTOUR}(b) and \ref{G_WEAK_CONTOUR}(d). In Fig. \ref%
{G_WEAK_CONTOUR}(d),\ the very narrow $W_{+}$ resonance peak at $\Delta
=-\Delta _{N}/2$ (corresponding to $B\approx 102.37$ mT, marked by white
dotted lines) and the $W_{-}$ resonance peak at $\Delta \approx \Delta
_{N}/2 $ (corresponding to $B\approx 102.45$ mT, marked by black dotted
lines) can be clearly resolved. In Fig. \ref{G_WEAK_CONTOUR}(b), the $W_{\pm
}$ resonances give rise to fine structures in the nuclear polarization $p_{%
\mathrm{ss}}$ near $B\approx 102.37$ mT and $B\approx 102.45$ mT,
superimposed on the usual dependence on the polar angle $\theta $ via $%
W_{\pm }\propto |A_{g,+\mp }|^{2}\propto (1\pm c_{\theta })^{2}$ [plotted on
top of Figs. \ref{G_WEAK_CONTOUR}(a) and \ref{G_WEAK_CONTOUR}(b)]. We notice
that a $^{13}$C nucleus located at $|\mathbf{R}|=20$ \AA\ away from the NV
center can be highly polarized with a DNP rate $\sim 60$ \textrm{Hz }$=(2.6\
\mathrm{ms})^{-1}$, sufficient to overcome the slow nuclear depolarization $%
\gamma _{\mathrm{dep}}\sim $ 1 s$^{-1}$.

Finally we present the three conditions for achieving strong nuclear
polarization:

\leftmargini=3mm

\begin{enumerate}
\item The linewidth $\Gamma\ll\left\vert \Delta_{N}\right\vert =|2\gamma
_{N}B-A_{g,zz}|$, where $2\gamma_{N}B\approx-2.2$ \textrm{MHz }since we
always consider $B\sim102$ mT near the GSLAC. Therefore, $\left\vert \Delta
_{N}\right\vert $ ranges from $\sim2$ MHz (dominated by $2\gamma_{N}B$) for
weakly coupled nuclei to $\sim100$ MHz (dominated by $A_{g,zz}$) for
strongly coupled nuclei.

\item For optimal performance, the $|-1_{g}\rangle $-$|0_{g}\rangle $
separation $\Delta $ should be tuned accurately to match one resonance peak.
This requires the mismatch $\delta \Delta <\Gamma ,|\Delta _{N}|$, i.e., the
magnetic field control precision $\delta B$ should be smaller than $|\Delta
_{N}|/\gamma _{e}$ ($\sim 0.8$ G for weakly coupled nuclei with $|\Delta
_{N}|\approx 2.2$ MHz near GSLAC) and $\Gamma /\gamma _{e}$ ($\sim 0.07$ G
for $\Gamma =0.2$ MHz), accessible by current experimental techniques, e.g.,
$\delta B=0.02$ G has been reported \cite{MaurerScience2012}. If the control
precision $\delta \Delta $ $\gg $ linewdith $\Gamma $, then the DNP rate
would be reduced by a factor $(\delta \Delta /\Gamma )^{2}$. Taking \ref%
{G_WEAK_CONTOUR}(d) as an example, a large magnetic field control error $%
\delta B=1$ G and hence $\delta \Delta =2.8$ MHz would reduce the maximal
DNP rate from $\sim 60\ \mathrm{Hz}$ to $\sim 2$ s$^{-1}$, which is still
comparable with the depolarization rate $\gamma _{\mathrm{dep}}\sim $ 1 s$%
^{-1}$.

\item For DNP to dominate over nuclear depolarization, the transverse HFI
strength should satisfy
\begin{equation*}
|A_{g,++}|^{2}+|A_{g,+-}|^{2}\gg 4\Gamma \gamma _{\mathrm{dep}},
\end{equation*}%
which reduces to Eq. (\ref{STRONG_HFI}) for dipolar HFI.
\end{enumerate}

\subsection*{DNP of multiple $^{13}$C nuclei}

Having established a completely picture for single-$^{13}$C DNP, now we
generalize the above results to many ($N\gg1$) $^{13}$C nuclear spins
coupled to the NV center via $\hat{H}_{\mathrm{HF}}=\sum_{i=1}^{N}\mathbf{%
\hat{F}}_{i}\cdot\hat{\mathbf{I}}_{i}$, where $\mathbf{\hat{F}}_{i}=\hat{%
\mathbf{S}}_{g}\cdot\mathbf{A}_{i}$ and the excited state HFI is neglected
because the induced NV-nuclei flip-flop processes are off-resonant. In the
same spirit as that in treating the DNP of a single nuclear spin, we
decompose the HFI into the longitudinal part $\sum_{i=1}^{N}\hat{F}_{i,z}%
\hat{I}_{i,z}\ $and the transverse part $\sum_{i=1}^{N}(\hat{F}_{i,+}\hat{I}%
_{i,-}+h.c.)/2$. Then we approximate the longitudinal HFI with $-\hat{\sigma}%
_{-1_{g},-1_{g}}\hat {h}_{z}$ (which is treated \textit{exactly}) and treat
the transverse HFI with second-order perturbation theory, where
\begin{equation*}
\hat{h}_{z}\equiv\sum_{i=1}^{N}A_{i,zz}\hat{I}_{i,z}
\end{equation*}
is the collective Overhauser field from all the nuclei.

The physics picture of the many-nuclei DNP is as follows. Up to leading
order, the flip of different nuclei by the transverse HFI is independent, in
the sense that at a given moment, only one nuclear spin is being flipped,
while other nuclear spins simply act as \textquotedblleft
spectators\textquotedblright . However, the flip of each individual nuclear
spin does depend on the states of all the nuclear spins via the collective
Overhauser field $\hat{h}_{z}$: each many-nuclei state $|\mathbf{m}\rangle
=\otimes _{i=1}^{N}|m_{i}\rangle $ ($|m_{i}\rangle $ is the Zeeman
eigenstate of the $i$th nucleus) produces an Overhauser field $h_{\mathbf{m}%
}\equiv \langle \mathbf{m}|\hat{h}_{z}|\mathbf{m}\rangle $ that shifts the
energy of the NV state $|-1_{g}\rangle $ by an amount $-h_{\mathbf{m}}$.
This renormalizes the $|-1_{g}\rangle $-$|0_{g}\rangle $ separation from $%
\Delta =D_{\mathrm{gs}}-\gamma _{e}B$ to $\Delta -h_{\mathbf{m}}$ and hence
changes the NV dynamics and NV-induced nuclear spin flip, e.g., the spin
flip rate $W_{i,\pm }(h_{\mathbf{m}})$ of the $i$th nucleus conditioned on
the many-nuclei state being $|\mathbf{m}\rangle $ is obtained from Eq. (\ref%
{WPM_5LEVEL_SPIN12}) by replacing $\mathbf{A}_{g},\Delta $ with $\mathbf{A}%
_{i}$, $\Delta -h_{\mathbf{m}}$, respectively.

The above physics picture is quantified by the following rate equations for
the diagonal elements $p_{\mathbf{m}}\equiv \langle \mathbf{m}|\hat{p}|%
\mathbf{m}\rangle $ of the many-nuclei density matrix $\hat{p}$:
\begin{align}
\dot{p}_{\mathbf{m}}& =-\sum_{i}\left[ W_{i,+}(h_{\mathbf{m}})p_{\mathbf{m}%
}-W_{i,-}(h_{\mathbf{m}+1_{i}})p_{\mathbf{m}+1_{i}}\right]  \notag \\
& \quad -\sum_{i}\left[ W_{i,-}(h_{\mathbf{m}})p_{\mathbf{m}}-W_{i,+}(h_{%
\mathbf{m}-1_{i}})p_{\mathbf{m}-1_{i}}\right] ,  \label{rateequation}
\end{align}%
which has been derived by adiabatically decoupling the fast electron
dynamics from the slow DNP process \cite%
{YangPRB2013,YangPRB2012,WangArxiv2015}. Here $|\mathbf{m}+1_{i}\rangle =%
\hat{I}_{i}^{+}|\mathbf{m}\rangle /\sqrt{\xi _{m_{i}}^{+}}$ is the same as $|%
\mathbf{m}\rangle $ except that the state of the $i$th nucleus changes from $%
|m_{i}\rangle $ to $|m_{i}+1\rangle $. Now we discuss the difference between
single-spin DNP and many-spin DNP. In the latter case, the DNP of each
individual nucleus occurs in the presence of many \textquotedblleft
spectator\textquotedblright\ nuclei, which produce a fluctuating Overhauser
field $\hat{h}_{z}$ that randomly shifts the NV energy levels, such that the
$|-1_{g}\rangle $-$|0_{g}\rangle $ separation changes from $\Delta $ to a
random value $\Delta -\hat{h}_{z}$. This makes it more difficult to tune the
magnetic field to match the $W_{+}$ (or $W_{-}$) resonance exactly. More
precisely, a finite mismatch $\sim (h_{z})_{\mathrm{rms}}$ makes the
originally resonant $W_{+}$ ($W_{-}$) process off-resonant, and hence reduce
the resonant DNP rate by a factor $(h_{z})_{\mathrm{rms}}^{2}/\Gamma ^{2}$.
For example, a natural abundance of $^{13}\mathrm{C}$ nuclei gives a typical
Overhauser field $(h_{z})_{\mathrm{rms}}\sim $ $0.2$ MHz. This reduces the
typical DNP rate by a factor of $2$ for the linewidth $\Gamma \approx R=0.2\
\mathrm{MHz}$, e.g., the maximal DNP rate $\sim 60\ \mathrm{Hz}$ [Fig. \ref%
{G_WEAK_CONTOUR}(d)] for a $^{13}$C nucleus at $2\ \mathrm{nm}$ away from
the NV center is reduced to $30$ $\mathrm{Hz}\approx (5\ \mathrm{ms})^{-1}$,
which is still sufficient to overcome the slow nuclear depolarization $%
\gamma _{\mathrm{dep}}\sim 1$ s$^{-1}$. Therefore, under weak pumping $%
\Gamma \ll |\Delta _{N}|$, the typical Overhauser field fluctuation $%
(h_{z})_{\mathrm{rms}}\ll |\Delta _{N}|$ does not significantly influence
the steady state nuclear polarization.

For $N$ nuclear spin-1/2's, the number of variables $\{p_{\mathbf{m}}\}$ is $%
2^{N}$. When $N$ is small, we can solve Eq. (\ref{rateequation}) exactly.
For large $N$, however, the exponentially growing complexity prohibit an
exact solution to Eq. (\ref{rateequation}). Here, we introduce an
approximate solution. The essential idea is to assume small inter-spin
correlation, so that the $N$-nuclei density matrix $\hat{p}$ can be
factorized as $\hat{p}=\otimes _{i=1}^{N}\hat{p}_{i}$, where the $i$th
nuclear spin state $\hat{p}_{i}\approx |\uparrow _{i}\rangle \langle
\uparrow _{i}|(1+p_{i})/2+|\downarrow _{i}\rangle \langle \downarrow
_{i}|(1-p_{i})/2$ is described by its polarization $p_{i}$. This
dramatically reduces the number of variables from $2^{N}$ many-nuclei
populations $\{p_{\mathbf{m}}\}$ to $N$ single-spin polarizations $\{p_{i}\}$%
. Substituting this approximation into Eq. (\ref{rateequation}) and tracing
over all the nuclei except for the $i$th nucleus give $N$ coupled equations:%
\begin{equation}
\dot{p}_{i}=-(\bar{W}_{i,+}+\bar{W}_{i,-})\left( p_{i}-\frac{\bar{W}_{i,+}-%
\bar{W}_{i,-}}{\bar{W}_{i,+}+\bar{W}_{i,-}}\right) ,  \label{recursion}
\end{equation}%
where $\bar{W}_{i,\pm }$ is the spin flip rate of the $i$th nucleus averaged
over the states of all the other $(N-1)$ nuclear spins:%
\begin{align*}
\bar{W}_{i,+}& =\mathrm{Tr}W_{i,+}(\hat{h}_{z})|\downarrow _{i}\rangle
\langle \downarrow _{i}|\otimes _{j(\neq i)}\hat{p}_{j}, \\
\bar{W}_{i,-}& =\mathrm{Tr}W_{i,-}(\hat{h}_{z})|\uparrow _{i}\rangle
\langle \uparrow _{i}|\otimes _{j(\neq i)}\hat{p}_{j},
\end{align*}%
i.e., $\bar{W}_{i,\pm }$ depends on the polarizations $\{p_{j}\}$ $(j\neq i)$
of all the other nuclear spins. Therefore, Eq. (\ref{recursion}) with $%
i=1,2,\cdots ,N$ form $N$ coupled differential equations for $\{p_{i}\}$.
The steady-state solutions $\{p_{i,\mathrm{ss}}\}$ are obtained by solving $%
N $ coupled nonlinear equations with recursive methods. This approach
provides a good approximation to the average polarization $\bar{p}_{\mathrm{%
ss}}=(1/N)\sum_{i=1}^{N}p_{i,\mathrm{ss}}$ and average Overhauser field $h_{%
\mathrm{ss}}=\sum_{i}A_{i,zz}p_{i,\mathrm{ss}}/2$, but does not include any
spin-spin correlation effect, such as feedback induced spin bath narrowing
\cite%
{GreilichScience2007,XuNature2009,VinkNatPhys2009,ToganNature2011,SunPRL2012,YangPRB2012,YangPRB2013,WangArxiv2015}%
.

\begin{figure}[tbp]
\includegraphics[width=\columnwidth]{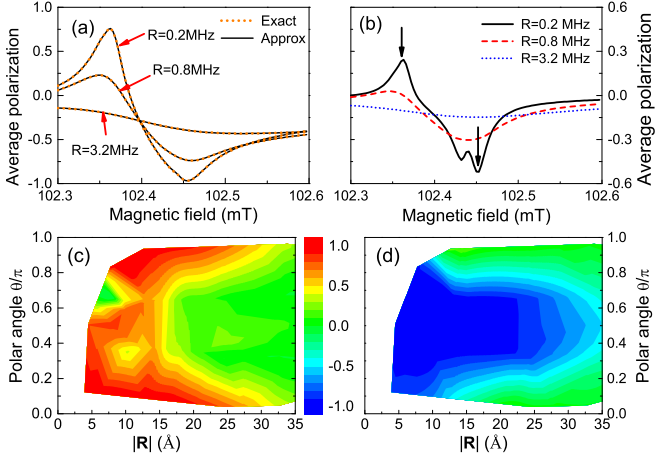}
\caption{Average polarization $\bar{p}_{\mathrm{ss}}$ of (a) $N=7$ and (b) $%
N=400$ randomly chosen $^{13}$C nuclei for different optical pumping rates $%
R=0.2,$ $0.8$, and $3.2$ MHz. For $R=0.2$ MHz, the spatial distribution of
the polarization of $N=400$ randomly chosen $^{13}$C nuclei in a magnetic
field (c) $102.36$ $\mathrm{mT}$ and (d) $102.45$ $\mathrm{mT}$ [marked by
the arrows in panel (b)]. The depolarization rate $\protect\gamma _{\mathrm{%
dep}}=1$ Hz for all panels.}
\label{G_MULTISPIN}
\end{figure}

As shown in Fig. \ref{G_MULTISPIN}(a), for a small number ($N=7$) of
randomly chosen $^{13}$C nuclei coupled to the NV via dipolar HFI, the
average polarization $\bar{p}_{\mathrm{ss}}$ from the above approximation
agrees well with the exact solution to the rate equations. For $N=400$
randomly chosen $^{13}$C nuclei, the exact solution is no longer available
and we plot the approximation results in Fig. \ref{G_MULTISPIN}(b). In both
Figs. \ref{G_MULTISPIN}(a) and \ref{G_MULTISPIN}(b), we can clearly see that
increasing the optical pumping strength significantly decreases the average
nuclear polarization and its magnetic field dependence due to the broadening
of $W_{\pm }$ resonances. For weak optical pumping $R=0.2$ MHz, the average
polarization $\bar{p}_{\mathrm{ss}}$ of $400$ $^{13}$C nuclei shows a
positive maximum $\sim +25\%$ at $B\approx 102.36$ mT (due to $W_{+}$
resonance) and a negative maximum $\sim -55\%$ at $B\approx 102.45$ mT (due
to $W_{-}$ resonance). The stronger polarization along $-z$ axis arises from
the dependence of the single-spin flip rate $W_{\pm }\propto |A_{g,+\mp
}|^{2}\propto (1\pm c_{\theta })^{2}$ on the polar angle $\theta $ of the
nuclear spin location: more nuclear spins have $|A_{g,++}|^{2}>|A_{g,+-}|^{2}
$ [as plotted on top of Figs. \ref{G_WEAK_CONTOUR}(a) and \ref%
{G_WEAK_CONTOUR}(b)] and hence favors negative polarization. These results
clearly demonstrate the possibility to strongly polarize weakly coupled $%
^{13}$C nuclear spins by using weak optical pumping near $W_{+}$ resonance ($%
B\approx 102.36$ mT) or $W_{-}$ resonance ($B\approx 102.45$ mT). The steady
state polarization of these weakly coupled $^{13}$C nuclei is ultimated
limited by nuclear spin depolarization. This can be clearly seen in the
spatial distribution of the nuclear polarization at the $W_{+}$ resonance
[Fig. \ref{G_WEAK_CONTOUR}(c)] and the $W_{-}$ resonance [Fig. \ref%
{G_WEAK_CONTOUR}(d)]. At $W_{+}$ resonance, strong positive polarization is
achieved for $^{13}$C nuclei with $|\mathbf{R}|<15$ \textrm{\r{A}} away from
the NV center, corresponding to dipolar HFI strength $>6$ kHz. For more
distant $^{13}$C nuclei, the HFI strength is too weak for the DNP to
dominate over the nuclear depolarization, so their polarization drops
significantly. Similarly, at $W_{-}$ resonance, strong negative polarization
can be achieved for $^{13}$C nuclei with $|\mathbf{R}|<25$ \textrm{\r{A}}
away from the NV center, corresponding to dipolar HFI strength $\sim 1$ kHz.
In principle, further decrease of the optical pumping rate $R$ and hence the
resonance linewidth $\Gamma $ could make the $W_{\pm }$ resonances even
sharper, which allows more distant $^{13}$C nuclei to be strongly polarized.

Finally, we recall that the on-site $^{15}$N or $^{14}$N nucleus has a
isotropic transverse HFI and hence $W_{-}=|A_{g,++}|^{2}=0$, i.e., they
could be completely polarized and thus does not significantly influence the
DNP of the $^{13}$C nuclei, except for a shift of the $|-1_{g}\rangle $-$%
|0_{g}\rangle $ separation $\Delta $ by $\sim 2.2$ MHz (for $^{14}$N) or $%
\sim 3$ MHz (for $^{15}$N), corresponding to a shift of the resonance
magnetic field by $\sim 0.78$ G (for $^{14}$N) or $\sim 1.1$ G (for $^{15}$%
N).

\section*{Discussion}

In conclusion, we have presented a comprehensive theoretical understanding
on the dynamic nuclear polarization induced by an optically pumped NV center
near the ground state anticrossing at ambient temperature. Our results not
only provide a clearly physics picture for a recently observed \cite%
{WangNatCommun2013} magnetic field dependence of the polarization of
first-shell $^{13}$C nuclei, but also reveals an efficient scheme to
strongly polarize weakly coupled $^{13}$C nuclear spins $\sim 25$ \AA\ away
from the NV center (HFI strength $\sim 1$ kHz) by weak optical pumping and
fine-tuning the magnetic field. These results provide a clear guidance for
optimizing future dynamic nuclear polarization experiments. For example,
this scheme could be used to polarize distant $^{13}$C nuclei in isotope
purified diamond \cite{BalasubramanianNatMater2009} to further prolong the
NV coherence time.

\section*{Methods}

\subsection*{Numerical evaluation of nuclear spin transition rates}

Taking $W_{m+1\leftarrow m}$ as an example, we need to first calculate $\hat{%
P}_{m}$ from $\mathcal{L}_{m,m}\hat{P}_{m}=0$ and $\mathrm{Tr}_{\mathrm{NV}}%
\hat{P}_{m}=1$, and then calculate $(\mathcal{L}_{m+1,m}-i\gamma_{N}B)^{-1}%
\hat{F}_{-}\hat{P}_{m}$. For this purpose, we map NV operators into vectors
and Liouville superoperators into matrices by introducing the complete basis
set $\{|ij)\equiv|i\rangle\langle j|\}$, where $i,j=1,2,\cdots,M$, and $M=7$
is the number of NV energy levels in our model [see Fig. \ref{G_ENERGYLEVEL}%
(a)]. Then the NV operator $\hat{F}_{-}\hat {P}_{m}=\sum_{ij}v_{ij}|ij)$ is
mapped to a vector $\mathbf{v}$ with components $v_{ij}=\langle i|\hat{F}_{-}%
\hat{P}_{m}|j\rangle$, and the Liouville superoperator $\mathcal{L}%
_{m+1,m}-i\gamma_{N}B$ is mapped to a matrix $\mathbf{S}$ via $(\mathcal{L}%
_{m+1,m}-i\gamma_{N}B)|ij)=\sum _{kl}|kl)S_{kl,ij}$, where the matrix element%
\begin{equation*}
S_{kl,ij}=\langle k|[(\mathcal{L}_{m+1,m}-i\gamma_{N}B)|i\rangle\langle
j|]|l\rangle.
\end{equation*}
Then we immediately obtain $(\mathcal{L}_{m+1,m}-i\gamma_{N}B)^{-1}\hat{F}%
_{-}\hat{P}_{m}=\sum_{ij}|ij)(\mathbf{S}^{-1}\mathbf{v})_{ij}$ and hence the
transition rates $W_{m+1\leftarrow m}$.

\subsection*{Explicit analytical expressions for nuclear spin transition
rates}

With $|+1_{g}\rangle$ and $|+1_{e}\rangle$ neglected, the rotating frame
Hamiltonian of the five-level NV model [Fig. \ref{G_ENERGYLEVEL}(b)] is%
\begin{equation*}
\hat{H}_{\mathrm{NV}}=\Delta\hat{\sigma}_{-1_{g},-1_{g}}+(D_{\mathrm{es}%
}-\gamma_{e}B+\omega_{0}-\omega)\hat{\sigma}_{-1_{e},-1_{e}}+\frac{\Omega_{R}%
}{2}(\hat{\sigma}_{-1_{e},-1_{g}}+h.c.).
\end{equation*}
Neglecting $\hat{F}_{z}$-induced NV spin mixing, the longitudinal HFI
reduces to $-(A_{g,zz}\hat{\sigma}_{-1_{g},-1_{g}}+A_{e,zz}\hat{\sigma}%
_{-1_{e},-1_{e}})\hat{I}_{z}$. Neglecting the non-collinear term $\propto%
\hat{S}_{z}\hat{I}_{\pm}$ (as the NV mostly stays in $|0_{g}\rangle$) and
the transverse NV excited state HFI (which is far off-resonant), the
transverse HFI reduces to $(\hat{F}_{+}\hat{I}_{-}+h.c.)/2$ with $\hat{F}%
_{+}=(\hat{\sigma}_{-1_{g},0_{g}}A_{g,++}+\hat{\sigma}%
_{0_{g},-1_{g}}A_{g,-+})/\sqrt{2}$. To calculate $W_{m\pm1\leftarrow m}$
from Eq. (\ref{WPM}), we first determine the NV steady state as $\hat{P}%
_{m,m}=P_{g}\hat{\sigma}_{0_{g},0_{g}}+(1-P_{g})\hat{\sigma}_{0_{e},0_{e}}$,
where $R=2\pi(\Omega_{R}/2)^{2}\delta^{((\gamma+\Gamma_{e}+\gamma_{%
\varphi})/2)}(\omega_{0}-\omega)$ is the optical pumping rate for $%
|0_{g}\rangle\leftrightarrow|0_{e}\rangle$. Substituting into Eq. (\ref{WPM}%
) gives%
\begin{equation}
W_{m\pm1\leftarrow m}=-\xi_{m}^{\pm}\frac{|A_{g,+\mp}|^{2}}{4}P_{g}\mathrm{Re%
}\rho_{-1_{g},0_{g}}^{\pm},  \label{A1}
\end{equation}
where $\rho_{i,j}^{(\pm)}\equiv\langle i|\hat{\rho}^{(\pm)}|j\rangle$ is the
$(i,j)$ matrix element of the operator $\hat{\rho}^{(\pm)}\equiv (\mathcal{L}%
_{m\pm1,m}\mp\gamma_{N}B)^{-1}\hat{\sigma}_{-1_{g},0_{g}}$, which is a
linear combination of $\hat{\sigma}_{-1_{g},0_{g}}$, $\hat{\sigma }%
_{-1_{e},0_{g}}$, $\hat{\sigma}_{-1_{g},0_{e}}$, and $\hat{\sigma}%
_{-1_{e},0_{e}}$ since the superoperator $\mathcal{L}_{m\pm1,m}$ for the
five-level NV model [Fig. \ref{G_ENERGYLEVEL}(b)] has no coherent coupling
between $\{|0_{g}\rangle,|0_{e}\rangle\}$ and $\{|-1_{g}\rangle,|-1_{e}%
\rangle\}$.

Now we calculate $\rho_{-1_{g},0_{g}}^{(\pm)}$ by taking the $\langle
0_{g}|\bullet|-1_{g}\rangle$,$\langle-1_{e}|\bullet|0_{g}\rangle$,$%
\langle-1_{g}|\bullet|0_{e}\rangle$ and $\langle-1_{e}|\bullet|0_{e}\rangle$
matrix elements of
\begin{equation*}
(\mathcal{L}_{m\pm1,m}\mp\gamma_{N}B)\hat{\rho}^{(\pm)}=\hat{\sigma}%
_{-1_{g},0_{g}},
\end{equation*}
which gives four coupled equations
\begin{align*}
&
(\Delta_{-1_{g},0_{g}}^{(\pm)}-i\Gamma_{-1_{g},0_{g}})\rho_{-1_{g},0_{g}}^{(%
\pm)}+\frac{\Omega_{R}}{2}(\rho_{-1_{e},0_{g}}^{(\pm)}-%
\rho_{-1_{g},0_{e}}^{(\pm)})=i, \\
&
(\Delta_{-1_{e},0_{g}}^{(\pm)}-i\Gamma_{-1_{e},0_{g}})\rho_{-1_{e},0_{g}}^{(%
\pm)}+\frac{\Omega_{R}}{2}(\rho_{-1_{g},0_{g}}^{(\pm)}-%
\rho_{-1_{e},0_{e}}^{(\pm)})=0, \\
&
(\Delta_{-1_{g},0_{e}}^{(\pm)}-i\Gamma_{-1_{g},0_{e}})\rho_{-1_{g},0_{e}}^{(%
\pm)}-\frac{\Omega_{R}}{2}(\rho_{-1_{g},0_{g}}^{(\pm)}-%
\rho_{-1_{e},0_{e}}^{(\pm)})=0, \\
&
(\Delta_{-1_{e},0_{e}}^{(\pm)}-i\Gamma_{-1_{e},0_{e}})\rho_{-1_{e},0_{e}}^{(%
\pm)}-\frac{\Omega_{R}}{2}(\rho_{-1_{e},0_{g}}^{(\pm)}-%
\rho_{-1_{g},0_{e}}^{(\pm)})=0.
\end{align*}
Here $\Delta_{j,i}^{(\pm)}$ is the energy difference between $|j\rangle
|m\pm1\rangle$ and $|i\rangle|m\rangle$ ($|i\rangle,|j\rangle$ are NV states
and $|m\rangle,|m\pm1\rangle$ are nuclear Zeeman states), i.e.,%
\begin{align*}
\Delta_{-1_{g},0_{g}}^{(\pm)} & =D_{\mathrm{gs}}-\gamma_{e}B\pm\gamma
_{N}B-(m\pm1)A_{g,zz}, \\
\Delta_{-1_{e},0_{e}}^{(\pm)} & =D_{\mathrm{es}}-\gamma_{e}B\pm\gamma
_{N}B-(m\pm1)A_{e,zz}, \\
\Delta_{-1_{e},0_{g}}^{(\pm)} & =E_{-1_{e},0_{e}}^{(\pm)}+\omega_{0}-\omega,
\\
\Delta_{-1_{g},0_{e}}^{(\pm)} & =E_{-1_{g},0_{g}}^{(\pm)}-\omega_{0}+\omega,
\end{align*}
and $\Gamma_{j,i}$ is the linewidth of the NV transition $|i\rangle
\leftrightarrow|j\rangle$, i.e., $\Gamma_{-1_{g},0_{g}}=\gamma_{\varphi}$, $%
\Gamma_{-1_{e},0_{e}}=\gamma+\gamma_{1}/2$, $\Gamma_{-1_{e},0_{g}}=(%
\Gamma_{e}+\gamma+\gamma_{1}+\gamma_{\varphi})/2$, and $%
\Gamma_{-1_{g},0_{e}}=(\Gamma_{e}+\gamma+\gamma_{\varphi})/2$. Eliminating $%
\rho _{-1_{e},0_{g}}^{(\pm)}$ and $\rho_{-1_{g},0_{e}}^{(\pm)}$ gives the
\textquotedblleft rate equations\textquotedblright:
\begin{align*}
& (\Delta_{-1_{g},0_{g}}^{(\pm)}-i\Gamma_{-1_{g},0_{g}}-\mathcal{R}^{(\pm
)})\rho_{-1_{g},0_{g}}^{(\pm)}+\mathcal{R}^{(\pm)}\rho_{-1_{e},0_{e}}^{(\pm
)}=i, \\
& (\Delta_{-1_{e},0_{e}}^{(\pm)}-i\Gamma_{-1_{e},0_{e}}-\mathcal{R}^{(\pm
)})\rho_{-1_{e},0_{e}}^{(\pm)}+\mathcal{R}^{(\pm)}\rho_{-1_{g},0_{g}}^{(\pm
)}=0,
\end{align*}
from which we obtain the solution%
\begin{equation}
\rho_{-1_{g},0_{g}}^{(\pm)}=\frac{i}{\Delta_{-1_{g},0_{g}}^{(\pm)}-i%
\Gamma_{-1_{g},0_{g}}-\mathcal{R}^{(\pm)}\left( 1+\frac{\mathcal{R}^{(\pm )}%
}{\Delta_{-1_{e},0_{e}}^{(\pm)}-i\Gamma_{-1_{e},0_{e}}-\mathcal{R}^{(\pm)}}%
\right) },  \label{RHOGG}
\end{equation}
where $\mathcal{R}^{(\pm)}=\mathcal{R}_{0}^{(\pm)}+\mathcal{R}_{-1}^{(\pm)}$
with $\mathcal{R}_{0}^{(\pm)}\equiv(\Omega_{R}/2)^{2}/(%
\Delta_{-1_{g},0_{e}}^{(\pm)}-i\Gamma_{-1_{g},0_{e}})$ and $\mathcal{R}%
_{-1}^{(\pm)}\equiv
(\Omega_{R}/2)^{2}/(\Delta_{-1_{e},0_{g}}^{(\pm)}-i\Gamma_{-1_{e},0_{g}})$
the \textit{complex} self-energy corrections to $|0_{g}\rangle$ and $%
|-1_{g}\rangle$ by the optical pumping $|0_{g}\rangle\leftrightarrow|0_{e}%
\rangle$ and $|-1_{g}\rangle\leftrightarrow|-1_{e}\rangle$, respectively.
More precisely, the optical pumping $|0_{g}\rangle\leftrightarrow|0_{e}%
\rangle$ ($|-1_{g}\rangle\leftrightarrow|-1_{e}\rangle$) induces an optical
Stark shift $\mathrm{Re}\mathcal{R}_{0}^{(\pm)}$ ($-\mathrm{Re}\mathcal{R}%
_{-1}^{(\pm)}$) and an effective dissipation $\mathrm{Im}\mathcal{R}%
_{0}^{(\pm)}$ ($\mathrm{Im}\mathcal{R}_{-1}^{(\pm)}$) for the NV ground
state $|0_{g}\rangle$ ($|-1_{g}\rangle$). Taking $\mathcal{R}_{0}^{(\pm)}$
as an example, if $|\Gamma_{-1_{g},0_{e}}|$ is much smaller than $%
|E_{-1_{g},0_{e}}^{(\pm)}|$, then the optical Stark shift $\delta
E_{0_{g}}\approx(\Omega_{R}/2)^{2}/\Delta_{-1_{g},0_{e}}^{(\pm)}$ reduces to
the conventional form of a second-order energy correction in non-degenerate
perturbation theroy, while the dissipation $\mathrm{Im}\mathcal{R}%
_{0}^{(\pm)}$ takes the semiclassical form of a Fermi golden rule.
Substituting Eq. (\ref{RHOGG}) into Eq. (\ref{A1}) immediately gives $%
W_{m\pm1\leftarrow m}$, which assumes a tedious form as it includes various
quantum coherence effects.

For simplification, we use the fact that the NV excited state dephasing $%
\Gamma_{e}\sim 10^{4}$ GHz $\gg $ other NV dissipation $\gamma ,\gamma
_{1},\gamma _{\varphi }$ and typical detuning $|\omega _{0}-\omega |$, $%
|\Delta _{-1_{g},0_{e}}^{(\pm )}|$, and $|\Delta _{-1_{e},0_{g}}^{(\pm )}|$,
and further restrict to weak optical pumping $R\ll \Gamma _{-1_{e},0_{e}}$ ($%
\sim 26.3$ MHz). In this case, the optical Stark shift is negligible and the
optical pumping rate simplifies to $R\approx \Omega _{R}^{2}/\Gamma_{e}$,
so the self-energies $\mathcal{R}^{(\pm )}\approx iR$ only induces NV level
broadening. Substituting the resulting expression $\rho
_{-1_{g},0_{g}}^{(\pm )}\approx i/(\Delta _{-1_{g},0_{g}}^{(\pm )}-i(\gamma
_{\varphi }+R))$ into Eq. (\ref{A1}) gives Eq. (\ref{WPM_5LEVEL}).


\section*{Acknowledgements}

This work was supported by NSFC (Grant No. 11274036 and No. 11322542) and
the MOST (Grant No. 2014CB848700).

\section*{Author contributions statement}

W. Y. and P. W. conceived the idea, formulated the theories, analyzed the
results, and wrote the paper. All authors discussed the results and the
manuscript.

\section*{Additional information}

Competing financial interests: The authors declare no competing financial
interests.

\end{document}